# Frequency Selective Surfaces for High Sensitivity Terahertz Sensing

Christian Debus, Peter Haring Bolivar

*Institut für Höchstfrequenztechnik und Quantenelektronik, Universität Siegen, Hölderlinstr. 3, 57076 Siegen, Germany*

An apporach for the sensing of small amounts of chemical and biochemical material is presented. A frequency selective surface (FSS) made from asymmetric split ring resonators (aDSR) is excited with free space radiation. Due to interference effects a resonance occurs with a steep flank in the frequency response which is shifted upon dielectric loading. Utilizing a strong E-field concentration by selective loading the detection of very small amounts of probe material becomes possible. The functionality is proven by numerical simulation and the optimization of structure and loading is performed.

Electromagnetic radiation in the THz frequency range has demonstrated an attractive potential for sensing chemical and biochemical compounds[1,2,3,4]. Sensing the complex dielectric properties of a sample in the THz frequency range enables the direct identification of chemical or biochemical molecular composition either by detecting the resonant absorption (i.e. imaginary part of the dielectric constant) of molecular or phonon resonances for small molecular compounds[3,4]. Or, for large biomolecules, as no sharp absorption features are observed, dielectric changes associated to the binding ("hybridization") of biomolecules are used for unambiguous detection and identification of macromolecules[1,2]. However, many envisaged applications of THz sensing systems, ranging from basic research instrumentation through biotechnology up to security relevant applications, require the detection of minute amounts of chemical and biomolecular substances. This is difficult to achieve with conventional THz spectroscopy systems given the huge difference between the sensing wavelength (several hundred micrometers) and the small nanometric scale of analyte quantity associated with typical applications. Flexible and sensitive solutions for probing the dielectric properties of minute quantities of chemical or biochemical compounds are therefore strongly desired. High sensitivity THz sensors are typically based on resonant structures whose frequency response is shifted by dielectric loading. Such a frequency shift can depend very sensitively on the dielectric properties of material placed in the environment of such a structure and can therefore be used to determine the dielectric constant of the sample material with high sensitivity. Several approaches have been demonstrated to this regard including microstrip resonators[5,6,7], Bragg reflection cavities[8], evanescent field resonances[9,10] and gratings[11]. Most of these approaches make use of planar waveguiding structures, leading to additional complications associated with the necessity to either integrate THz sources and detectors in the vicinity of the sensing elements, or requiring complicated or bulky THz coupling structures. High sensitivity resonant sensors which are directly accessible from free space are therefore highly desired. This minimizes system complexity and allows the implementation of separate reader and sensor chips systems, as sensors in many cases have to be conceived as throw-away articles to prevent cross contamination between samples. In this paper we propose the use of frequency selective surfaces (FSS) as a THz sensing approach which can directly be accessed from free space. Numerical calculations of the behaviour of FSS consisting of asymmetric double split ring (aDSR) elements are presented showing a very promising sensitivity.

To achieve a high sensitivity a sensor needs to have a sharp edge in its frequency response to enable the detection of small changes in the dielectric environment (sensitivity increases with the steepness of spectral features) plus the electric field distribution needs a point of high concentration to enable sensing small amounts of probe material. Split ring resonators used for metamaterials show high E-field concentrations[12]. They have also been arrayed to form FSSs[13] and analysed as a passive structure with THz spectroscopy[14]. By adding a second gap to such split rings and breaking the symmetry additional modes to the base mode can be excited, as demonstrated in theory in other frequency ranges[15]. However, the simplicity of such a FSS structure enables the utilization at any frequency by simple rescaling. Manufacturing with standard semiconductor technologies provides good characteristics for FSS at terahertz frequencies[16,17]. Manufacturing is simple, e.g. sputtering thin metal layers on dielectric or silicon substrates and structuring them in a single photolithography process and etching, enabling cheap mass production of FSS based THz sensors.

In our approach metallic structures with a thickness of 400 nm and a width of w = 5 µm are used. FIG. 1 depicts a section of an aDSR FSS and the dimensioning of each unit cell. With an asymmetry angle of $\varphi = 0°$ the DSR is symmetric to both x- and y-axis. Increasing the angle $\varphi > 0°$ shifts both gaps in positive x-direction, creating an asymmetry with respect to the y-axis.

Numerical calculations are performed using a commercial 3-D finite element method solver (HFSS by Ansoft Corp.). A single unit cell and periodic boundary conditions in x- and y-direction are adopted, so that the FSS is assumed to be infinitely large. A plane wave excitation propagating normally to the xy-plane and polarized in y-direction is assumed. Material properties for gold are computed with the Drude-Model.

The reflection of the FSS shows two significant features (FIG. 2). A broad maximum at 1090 GHz is observed, where the length of each arc of the aDSR approximately matches half the wavelength. This dipole antenna like behavior shows up for the symmetric DSR ($\varphi = 0°$) and is not significantly affected by small $\varphi$ angles. By increasing the angle $\varphi > 0°$, the two arcs of the DSR become differently long. Around 875 GHz the reflection of this asymmetric case shows a strong and sharp modulation of 13 dB over 13 GHz. Outside the 3-dB-ranges a flank with a very high steepness of 7 dB over 4 GHz is maintained for a FSS made from gold. It is interesting to observe that at this steep flank the electric field concentrates strongly close to the ring with amplitudes 25 times higher than the excitation $E_{inc}$ (FIG. 3).

Like in an antenna the electric field of the plane wave excitation causes a primary current $A_{l/r}$ along each arc. In the asymmetric case the left (l) and right (r) arc have slightly different lengths and therefore slightly different resonance frequencies and currents. Each primary current causes a secondary E-field which couples to the other arc and excites a secondary current $-K_{r/l}$ directed opposite to $A_{l/r}$. The sum $I_{l/r}$ of the currents defines the strength of an arc's resonance:

$$I_l = A_l - K_r \quad (1)$$

$$I_r = A_r - K_l \quad (2)$$

At 1090 GHz, the spectral response is dominated by the primary excitation, the base mode is excited. Towards lower frequencies the primary excitation increasingly looses strength. At 885 GHz the coupling between the arcs becomes identical to the primary excitation in both directions, $K_r = A_l$ and $K_l = A_r$, from which $I_l = I_r = 0$ follows. Without currents the field can pass the structure unhindered and the frequency response shows a total transmission for the lossless case (FIG. 2, dottet line). At a slightly higher wavelength the current $I_l$ in the longer arc reaches a maximum and the current $-K_l$ in the shorter arc becomes maximal, so that $I_r$ is directed opposite to the primary excitation. Being resonant in this mode the reflection of the aDSR shows a local maximum (867 GHz). Towards higher frequencies the corresponding antisymmetric current mode shows a less distinct reflection maximum and merges into the broadband reflection. Losses inside the conductor restrain the extrema (FIG. 2, solid line) but maintain the basic spectral shape of this dual resonance feature (DRF).

The main reason for the DRF is the interference effect associated with the small difference in the length of the DSR arcs. Changing the size dφ of the gaps shifts the DRF but does not significantly affect the shape as the length difference remains stable. However, changing the asymmetry angle φ varies the arc lengths difference and therefore the frequency distance between the interfering resonances (FIG. 4a). While the DRF shows a steep flank with $\varphi = 4°$ it becomes smaller and less steep towards smaller angles. Towards larger angles the modulation increases, but the flank becomes significantly flatter. The steepness of the DRF flank is analyzed by evaluating the frequency derivative of the spectra as depicted in FIG. 4 (b) and (c). The results indicate a maximum steepness at $\varphi = 4°$.

Finally, it is interesting to analyze the sensitivity of the optimized FSS structures for sensing applications, i.e. the shift of the resonance towards lower frequencies upon loading with small amounts of dielectric material. An analyte with a dielectric constant typical for organic systems and a film thickness of 10nm which corresponds to typical monomolecular film thicknesses in biosensing applications is assumed. With the complete surface covered with sample material, the DRF is shifted 5 GHz (FIG. 5a). In view of maximizing the sensitivity, much smaller sample coverings are explored, e.g. with a 27.5 µm x 27.5 µm area covering. Placing such a localized sample in the middle of the cell leads to a negligible effect ($\Delta f < 1$ MHz) on the DRF frequency, since the E-field strength at this location is almost zero. However, placing the sample in a spot with a high E-field concentration (FIG. 5b) maximizes the effect. The DRF frequency shifts by 4 GHz, which is almost as strong as for the full covering case. A large and easily detectable change in the THz reflection amplitude of 6 dB at 873°GHz is obtained with 1/64 of the amount of material of the full covering case.

Other approaches with stripline-shape resonators of comparable size require significantly larger amounts of probe material. Simulations of a ring resonator[18] do not show a distinct frequency shift with a loading of less than 80 nm thickness. Such an approach would require an 800 times thicker layer of probe material ($\varepsilon = 1.44$) to achieve the same shift of that of the FSS. Applying filters to amplify the frequency shift leads to smaller spots of material[7], but still the material volume requirement is approximately 1000 times higher to a achieve a shift comparable to the FSS calculation shown above. The corrugated waveguide approach[6] features smaller frequency shifts of less than 2 GHz at 1 µm thickness ($\varepsilon = 2.6$). However, alike the FSS this chip shows disproportionate strong shifting with selective loading compared to full layer loading.

A promising concept for high sensitivity THz chemical and biochemical sensors is presented. FSS can directly be read out from a free space geometry, enabling simple sensor system realizations. Frequency selective surfaces based on asymmetric double split ring structures exhibit two crucial features needed for high sensitivity sensing: a steep spectral feature to detect small frequency shifts reliably and a high E-field concentration for sensing small amounts of probe material. A dual resonance interference feature is observed with a steep flank of 7 dB over 4 GHz. This feature can significantly be shifted at small analyte loading densities, e.g. 4 GHz by covering selectively only 1/64 of the surface with a 10nm thick organic material. The sensing potential of this concept in combination with a localized sensor functionalization technology can therefore be regarded as high. We gratefully acknowledge financial support by the European Commission and by the Deutsche Forschungsgemeinschaft.

FIG. 1 (a) Schematic section of aDSR based FSS adopting a square lattice. (b) Unit cell with: radius $r = 50$ µm, width $w = 5$ µm, cell size $cs = 220$ µm, asymmetry angle $\varphi = 4°$ and gap angle $d\varphi = 20°$.

FIG. 2 Reflection of FSS of symmetric (dashed line) and asymmetric DSRs with $\varphi = 4°$ for a perfect conductor (dotted line) and for gold (solid line).

FIG. 3 The E-Field in the resonator plane shows a strong concentration (white) at the ends of the arcs. $f = 875$ GHz, amplitude of excitation 1 V/m.

FIG. 4 Dependency of the reflection on the asymmetry angle $\varphi$ depicting (a) reflection spectra, (b) the "steepness" = frequency derivative of the reflection amplitude, and (c) the maximum steepness as a function of the gap position.

FIG. 5 DRF frequency shift upon dielectric loading ($\varepsilon = 3.2$) with a 10nm thick film. (a) Loading covers the complete surface. (b) Partial loading (27.5 x 27.5 $\mu m^2$ area) at the position of maximum E-Field concentration.

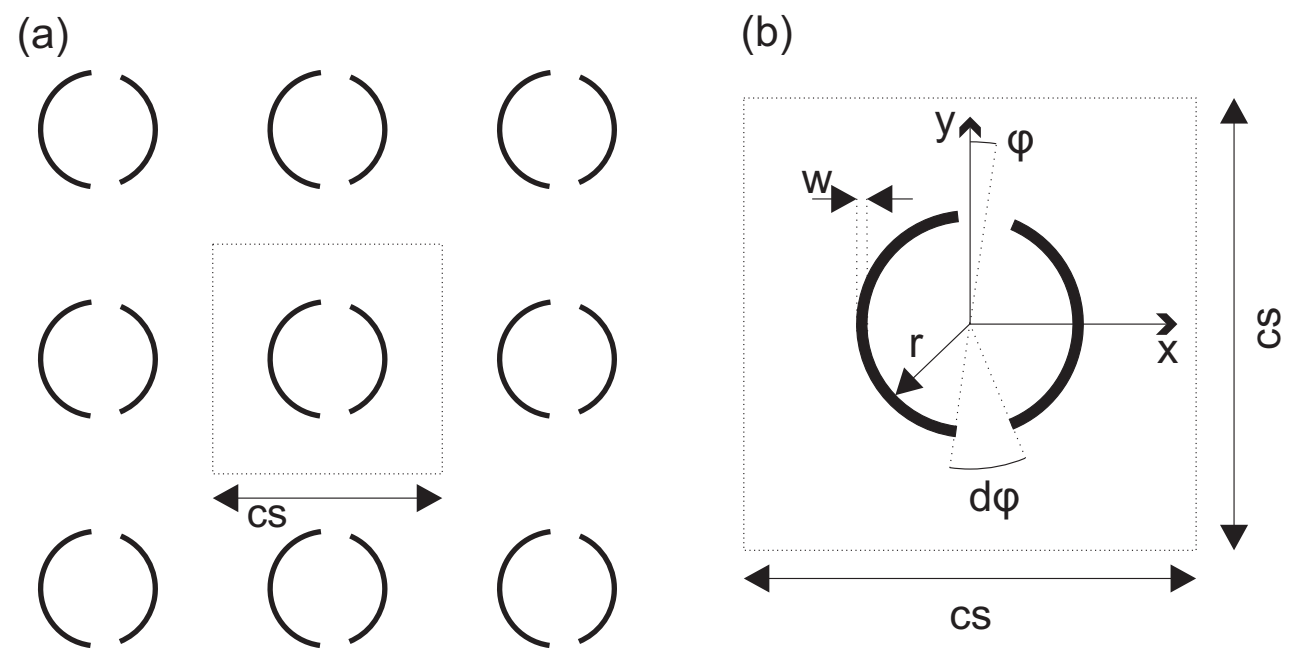

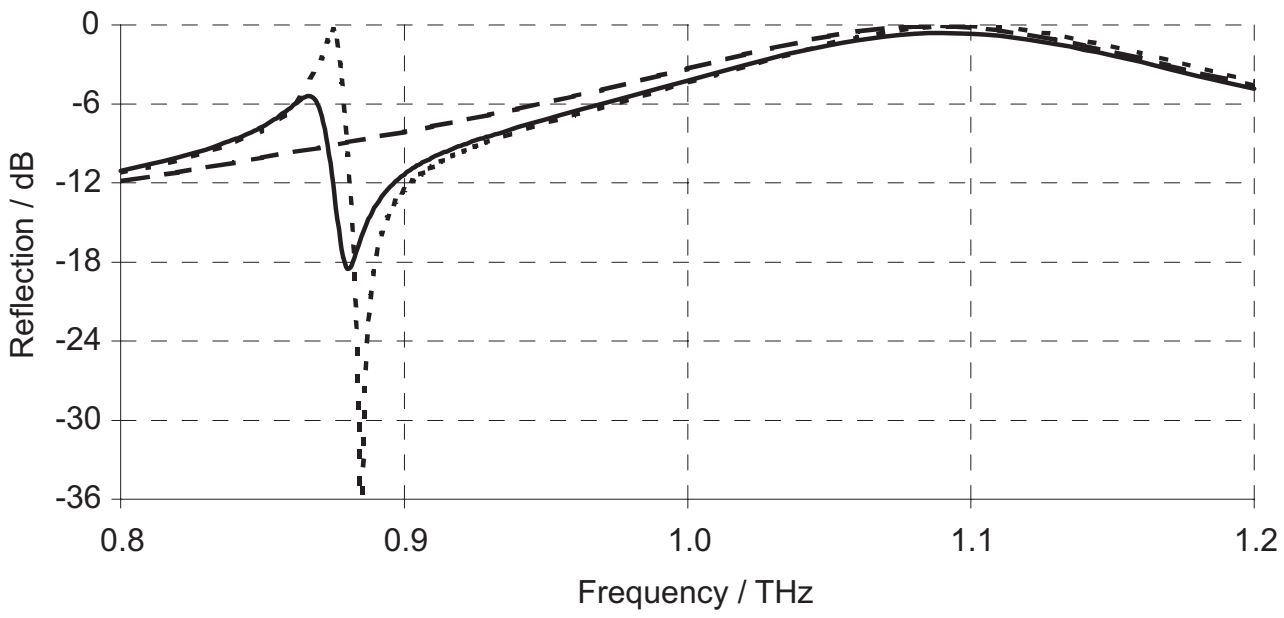

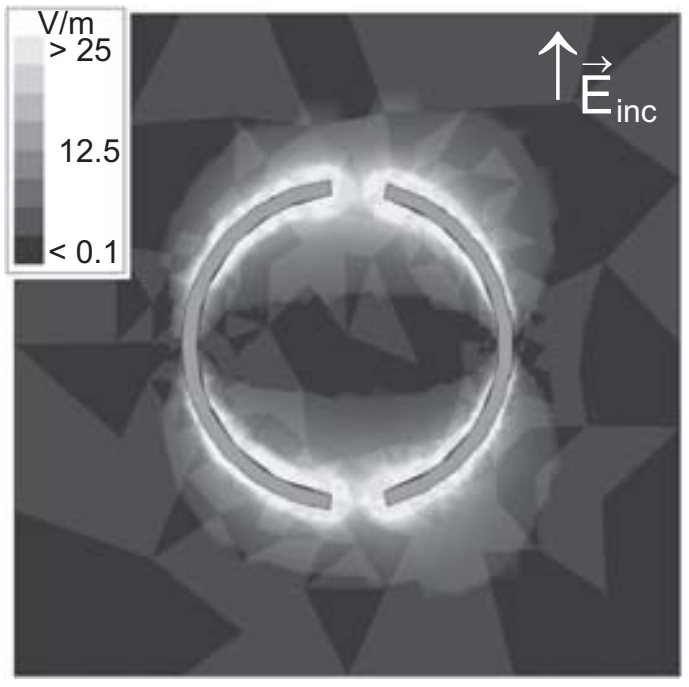
V/m
> 25
12.5
< 0.1
$\vec{E}_{inc}$

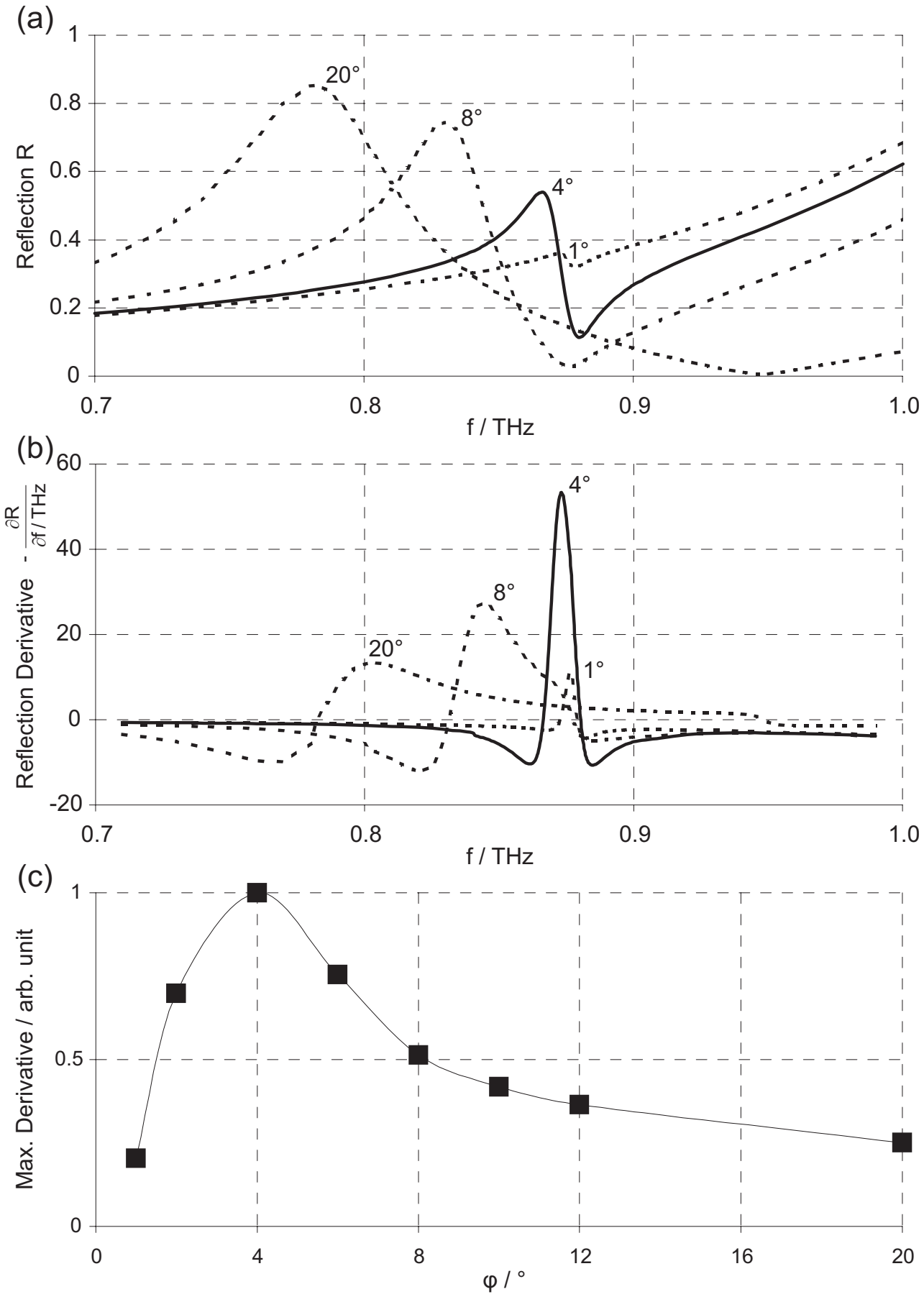
(a)
Reflection R
1
0.8
0.6
0.4
0.2
0
0.7
0.8
0.9
1.0
f / THz
20°
8°
4°
1°
(b)
Reflection Derivative $-\frac{\partial R}{\partial f / THz}$
60
40
20
0
-20
0.7
0.8
0.9
1.0
f / THz
4°
8°
20°
1°
(c)
Max. Derivative / arb. unit
1
0.5
0
0
4
8
12
16
20
φ / °

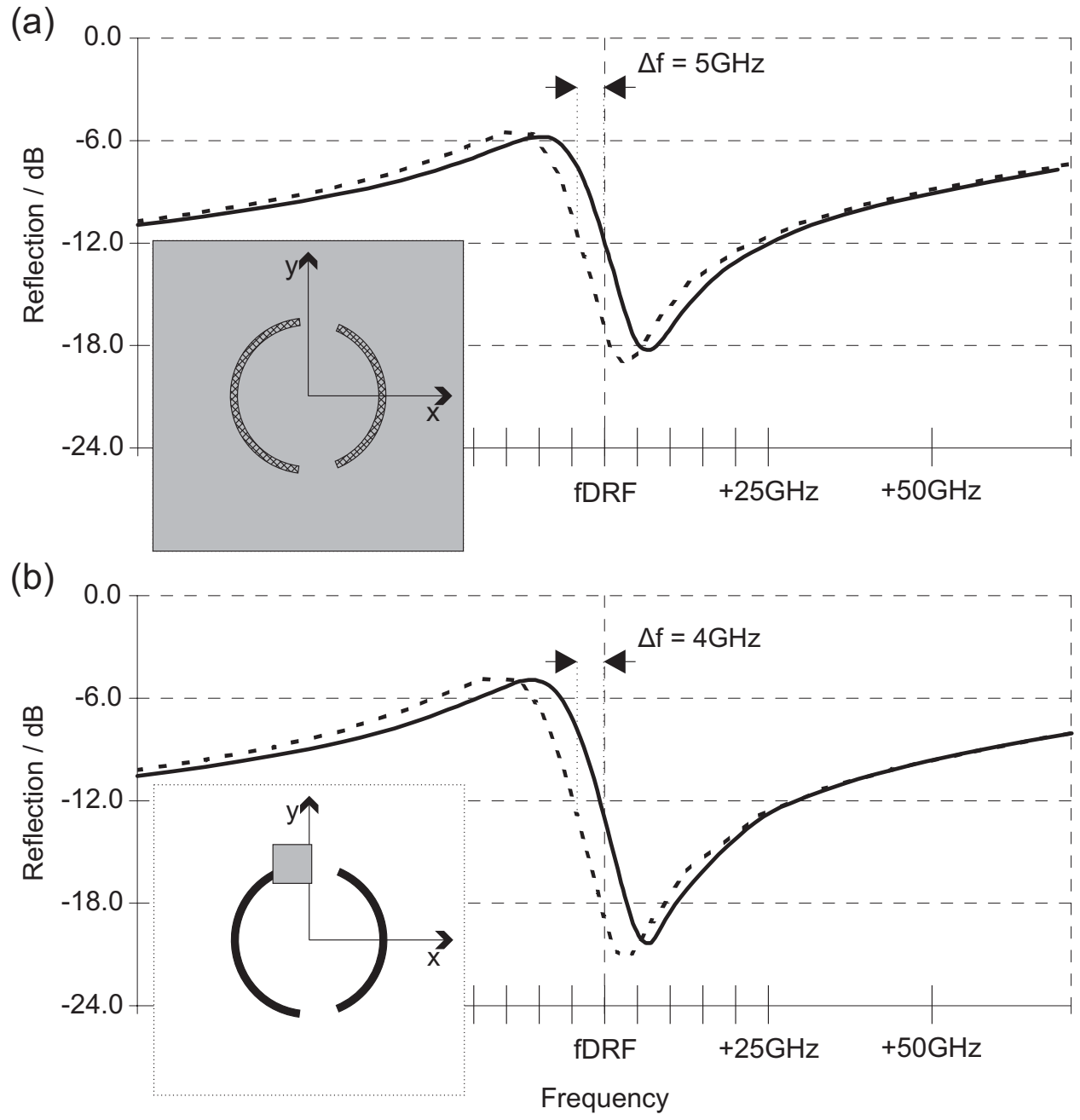